\documentclass[final,5p]{elsarticle}

\usepackage{amssymb}
\usepackage{subcaption}
\usepackage{amsthm}
\usepackage[T2A]{fontenc}			
\usepackage[utf8]{inputenc}			
\usepackage{amsfonts,amssymb,amsthm,mathtools} 
\usepackage{amsmath}
\usepackage{color}

\journal{Journal of Molecular Liquids}

\begin{document}

\begin{frontmatter}



\title{Can we accurately calculate viscosity in multicomponent metallic melts?}


\author[inst1,inst2,inst3]{Nikolay Kondratyuk}
\author[inst4,inst5,inst6]{Roman Ryltsev}
\author[inst4]{Vladimir Ankudinov}
\author[inst4]{Nikolay Chtchelkatchev}

\affiliation[inst1]{organization={Moscow Institute of Physics and Technology (National Research University)},
            addressline={Institutskiy Pereulok 9}, 
            city={Dolgoprudny},
            postcode={141701}, 
            state={Moscow oblast},
            country={Russia}}

\affiliation[inst2]{organization={Joint Institute for High Temperatures, Russian Academy of Sciences},
            addressline={Izhorskaya 13 Bldg 2}, 
            city={Moscow},
            postcode={125412}, 
            country={Russia}}
            
\affiliation[inst3]{organization={Higher School of Economics (National Research University)},
            addressline={Myasnitskaya 20}, 
            city={Moscow},
            postcode={101000}, 
            country={Russia}}

\affiliation[inst4]{organization={Vereshchagin Institute of High Pressure Physics, Russian Academy of Sciences},
            addressline={Kaluzhskoe sh. 14}, 
            city={Moscow (Troitsk)},
            postcode={108840}, 
            country={Russia}}

\affiliation[inst5]{organization={Institute of Metallurgy of the Ural Branch of the Russian Academy of Sciences},
            addressline={Amundsen str. 106}, 
            city={Ekaterinburg},
            postcode={620016}, 
            country={Russia}}
            
\affiliation[inst6]{organization={Ural Federal University},
            addressline={Lenin Ave, 51}, 
            city={Ekaterinburg},
            postcode={620002}, 
            country={Russia}}

\cortext[cor1]{Corresponding address: kondratyuk@phystech.edu}

\begin{abstract}
Calculating viscosity in multicompoinent metallic melts is a challenging task for both classical and \textit{ab~initio} molecular dynamics simulations methods. The former may not to provide enough accuracy and the latter is too resources demanding. Machine learning potentials provide optimal balance between accuracy and computational efficiency and so seem very promising to solve this problem. Here we address simulating kinematic viscosity in ternary Al-Cu-Ni melts with using deep neural network potentials (DP) as implemented in the DeePMD-kit. We calculate both concentration and temperature dependencies of kinematic viscosity in Al-Cu-Ni and conclude that the developed potential allows one to simulate viscosity with high accuracy; the deviation from experimental data does not exceed 9\% and is close to the uncertainty interval of experimental data. More importantly, our simulations reproduce minimum on concentration dependency of the viscosity at the eutectic point. Thus, we conclude that DP-based MD simulations is highly promising way to calculate viscosity in multicomponent metallic melts.
\end{abstract}

\begin{graphicalabstract}
\includegraphics{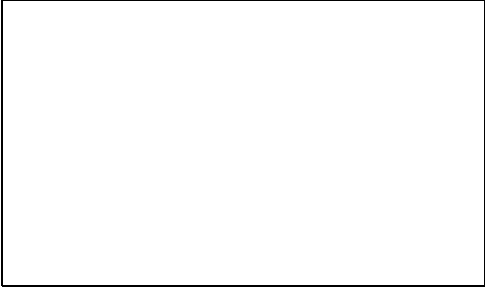}
\end{graphicalabstract}

\begin{highlights}
\item Deep machine learning potential is applied for simulating viscosity of Al-Cu-Ni melts
\item Results match experimental viscosity dependencies on temperature and concentration
\item Simulations reproduce the minimum of the viscosity at the eutectic composition
\item Deep machine learning potentials predict viscosity of multicomponent metallic melts
\end{highlights}

\begin{keyword}
viscosity \sep molecular dynamics \sep machine learning potentials \sep metallic melts 
\end{keyword}

\end{frontmatter}


\section{INTRODUCTION}
\label{sec:intro}

An accurate calculation of transport coefficients in liquids, such as shear viscosity and diffusivity, is one of the most important tasks in condensed matter theory. Indeed, the characteristics of atomic transport are crucial for studying such fundamental processes as glass formation and nucleation as well as for constructing and verifying model theories of liquid state~\cite{Archer2006} and coarse grained phase-field models~\cite{gdl-grant2008,jgrs-finel-08,jg13,Ankudinov2020a}. From a practical point of view, the values of the transport coefficients of melts influence the performance  of metallurgical casting and other pyrometallurgical processes, the creation of heat transfer fluids, the improvement of ion batteries, and the development of new lubricants, etc.~\cite{Meier2017}.

A reliable experimental determination of the viscosity and the diffusivity is a difficult and sometimes nearly impossible task. Among common experimental difficulties are (i) high temperatures and pressures that can be hardly possible to achieve in the experiment (e.g. in silicate melts)~\cite{Huaiwei2015RevGeophys}; (ii) strong chemical interaction of the melts with the environment and sample containers (e.g. for Al-based and Zr-based melts); (iii) low accuracy and/or high cost of certain experimental techniques (e.g., measuring of the  diffusion coefficients)~\cite{Masaki2005MeasurSciTech,Meyer2015EPJWeb}. In this regard, atomistic computer simulations acquire a special role due to the possibility to calculate any characteristics of atomic transport based on the trajectories of particles. Within the molecular dynamics methods, the shear viscosity coefficients could be obtained via the equilibrium (Green--Kubo~\cite{Green1954,Kubo1957}, Helfand~\cite{Helfand1960}, finite size effects of diffusion~\cite{Jamali2018,Orekhov2021}) or non-eqilibrium methods (M\"{u}ller-Plathe~\cite{Bordat2002}, SLLOD~\cite{Mundy1995JCP}). At the same time, the methods for viscosity calculation are shown to be quite accurate for various systems from atomic~\cite{kuksin2010theory,Fomin2012jetp,Rudyak2014,Kim2018JCP,Heyes2019,Bell2020,kirova2020dynamics} to molecular fluids~\cite{Glova2019,Camp2020,Bakulin2021,Ewen2021,Mathas2021,Yang2022,Deshchenya2022}. In theory, the predictive power of such models depends only on the accuracy of the predefined interatomic forces. These forces could be described via the model potentials, Machine Learning Interatomic Potentials (MLIP) and Density-Functional Theory (DFT), which are listed in ascending order of the required computational resources. The key factor of the choice among this approaches is the balance between accuracy and computational efficiency~\cite{Stegailov2019,Stegailov2021,Shaw2021}. 

Traditional methods based on classical molecular dynamics (CMD) operate with relatively simple model potentials and allow one to simulate systems of ${10^6-10^9}$ atoms at times up to microseconds. These spatio-temporal scales are sufficient to obtain the shear viscosity coefficients via both equilibrium and non-equilibrium methods. For example, CMD with effective pair potentials resulted to the viscosity of liquid Li, Na and Rb in close agreement with experimental data~\cite{Canales1994, Meyer2016, Demmel2018}. The embedded atom model (EAM)~\cite{Daw1984,Finnis1984} is a wide spread multibody potential for the CMD simulations of metals. For liquid Na and Al, EAM potentials provide a relatively good accuracy for the calculated viscosity in comparison to the experiment~\cite{Metya2012,Kirova2019,kirova2020dynamics}. On the other hand, the results for viscosity of Li~\cite{Metya2012} and Ni~\cite{Cherne1998} melts obtained from EAM-based MD simulations deviate from experimental values by a factor of two. The review~\cite{Cherne2001} shows that different functional forms of  liquid Ni potential lead to the  significantly differing values of viscosity. Moreover Mendelev et al.~\cite{Mendelev2008}  demonstrate that the potentials, which are parameterized for the properties of the crystalline phase only, poorly describe the liquid phase. 

Thus, the accuracy of CMD is very much limited by using of the empirical potentials, which in many cases are unable to adequately approximate the complex nature of interatomic interactions in real systems. This is especially evident in the case of multicomponent metallic melts. 


Meanwhile, highly accurate \emph{ab~initio} methods are very resource-demanding and therefore allow simulating physical systems only at spatio-temporal scales, which are insufficient for reliable calculation of viscosity.  The crucial problem is the requirement of the relatively  long trajectories   to perform time averaging for the viscosity, since it is a collective property. These time scales are still very computationally expensive and hardly achievable by DFT. A possible solution is to use indirect diffusivity-based methods like the Stokes-Einstein (S-E) relation where the diffusivity multiplied by the viscosity is proportional to the temperature of the system and the inverse hydrodynamic radius of a particle. The methods for calculation of the  diffusion constants use the ensemble averaging in addition to the  time averaging, and therefore are achievable within DFT method. This method of the shear viscosity estimations was applied to Al$_{80}$Cu$_{20}$~\cite{Wang2015}, GeTe~\cite{Weber2017}, and  NaCl-CaCl$_2$~\cite{Rong2021}. Adjaoud et al.~\cite{Adjaoud2011} show the applicability of the S-E relation for liquid Mg$_{2}$SiO$_{4}$ in a wide range of pressures and temperatures. However, the problem with the S-E method arises when considering multi-component systems: the diffusivity becomes a tensor and it stays unclear which component of the tensor should be used for the estimations.


Thus, we conclude that accurate calculation of the shear viscosity in multicomponent metallic melts is a challenging task because it requires both high accuracy and high computational efficiency. Fortunately, a machine-learning-based approach has recently emerged that makes it possible to effectively solve this problem~\cite{Ceriotti2021JCP,Mishin2021ActaMater,VonLilienfeld2020NatComm,Behler2021EPJ,Deringer2020JPhysEnerg,Mueller2020JCP,Deringer2019AdvMater,Behler2016JCP}. The idea is to approximate the potential energy surface of the system by some general many-body function (for example, a multilayer neural network) using reference \textit{ab~initio} data. Such machine learning interatomic potentials (MLIPs) can provide a nearly \textit{ab~initio} accuracy with several orders of magnitude less computational costs~\cite{Zuo2020JPhysChemA}. In comparison with traditional analytical MD potentials, usage of MLIPs is especially beneficial for the systems with highly diverse local atomic environment: e.g. for high-entropy metal alloys~\cite{kostiuchenko2019impact} or for multiphase carbon systems~\cite{nguyen2021billion,orekhov2022atomistic}.

Despite high potential of using MLIPs to calculate viscosity in metallic melts, there are only few examples of such application in published literature for the best of our knowledge. Lopanitsyna et al.~\cite{Lopanitsyna2021} developed such a potential for liquid Ni and obtain the temperature dependence of the viscosity via finite size effects of diffusion. A reasonable agreement with experimental data was achieved. Another example, is a successive use of deep machine learning potential for calculating viscosity of liquid gallium~\cite{Balyakin2022CompMaterSci}. The mentioned papers address viscosity of pure metals; as far as we know, similar works devoted to even binary metallic melts are absent. The authors familiar with only a recent successful application of MLIPs for the viscosity of molten binary salts~\cite{Feng2022JML}. The possible reasons are the relative novelty of MLIP-based simulations and the lack of reliable experimental data on the viscosity of multi-component melts. 

Here we make a step to fill this gap and apply MLIP-based simulations for calculating shear viscosity in the ternary Al-Cu-Ni system. This system demonstrates complicated chemical interaction between its components, which leads to a non-monotonous behaviour of the kinematic viscosity as a function of a concentration~\cite{Kamaeva2020}.


\section{Neural network potential}

\subsection{Training procedure \label{sec:train}}

\begin{table*}
\caption{Details of networks structure and training scheme used to generate DPs. The vectors determining the structure of both embedding and fitting nets contain the numbers of the neurons in each network layer. The cutoffs vector is $(r_{\rm cs}, r_{\rm c})$, where $r_{\rm cs}$ is the smooth cutoff parameter and $r_{\rm c}$ is the cutoff radius of the model. The learning scheme vectors contain the start and end values of loss function contributions  from energies, forces and virial tensors, respectively.}\label{tab:model_parameters}
\centering
\begin{tabular}{  c  c | c  c  | c   }
\hline
 embedding net  & axis neuron & fitting net  & cutoffs & learning  scheme  \\
 \noalign{\smallskip}

\hline
\noalign{\smallskip}

(25, 50, 100)  & 8 &    (240, 240, 240)      &(1.5,7)& (0.1, 1.0); (1000, 1.0); (0.05, 0.1) \\
    
\end{tabular}
\end{table*}

As a tool for developing MLIPs, we use DeePMD-kit, which utilizes feedforward multilayer neural networks to build highly accurate and efficient interatomic potentials (hereafter, Deep Potentials (DP))~\cite{Wen2022MaterFut}. Among other similar approaches, DeePMD uses an effective and automatic procedure for mapping the particles' coordinates to the space of structural descriptors, which includes many-body interactions and provides invariance with respect to translations, rotations, and permutations. Another advantage is the DPGEN tool, which utilizes concurrent learning strategy~\cite{Zhang2020CompPhysComm,Zhang2019PRM} for generating compact and representative training datasets and creating DPs, which are uniformly accurate in the whole space of thermodynamic parameters. This approach has been successfully applied for simulating systems of different nature~\cite{Wen2022MaterFut,Ryltsev2022JMolLiq,Niu2020NatComm,Sommers2020PCCP,Gartner2020PNAS,Balyakin2020PRE,Wen2019PRB,Tang2020PCCP,Zhang2021PRL,Andolina2020JCP}.

The training dataset for DP's developing  was generated within the framework of concurrent learning strategy using DPGEN package by the following scheme. We considered 14 compositions along the Al$_{100-x}$Cu$_{x}$Ni$_{10}$ cross-section where $x$ varied in range of $17-40$ at.\%. For each composition, we generate by {\it ab~initio} molecular dynamics simulations a supercell of 512 atoms which structure corresponds to the equilibrium melt. These configurations has been applied as initial ones for concurrent learning procedure.

Concurrent learning as implemented in DeePMD and DPGEN packages is an iterative procedure, which consists of three repeating stages. At the first (training) stage, we use the dataset to train an ensemble of four DPs with different initializations of neural network weights. As the initial dataset we used several tens of configurations and corresponding values of energies, forces and virial tensors extracted from {\it ab~initio} molecular dynamics simulations. The parameters used for the training are presented in Tab.~\ref{tab:model_parameters}. Each model was trained for 400,000 iterations with the batch size equal to 1.  

At the second (exploration) stage, we use one of the DPs obtained from the ensemble to perform CMD simulations of 14 Al$_{100-x}$Cu$_{x}$Ni$_{10}$ alloys at temperature $T=1600$ K, which corresponds to equilibrium liquid for all compositions under consideration. The length of each MD run was 25 ps (25,000 steps of 1 fs). At each MD step, we perform analysis of resulting atomic configurations and select candidates for DFT labelling and including to an extended training dataset. The selection criterion is the value of the model deviation parameter $\epsilon$, which is the maximal deviation of the forces predicted by the ensemble of models. The idea behind this parameter is very simple. For 'good' configurations, which are reliably described by a DP, $\epsilon$ is expected to be small. If $\epsilon$ is rather large, then corresponding configuration probably belongs to the extrapolation region for current dataset (not covered by the training data) and thus should be treated as a candidate for DFT labelling. In practice, one should determine two parameters: upper $\sigma_{up}$, and lower $\sigma_{lo}$ trust levels, which are used to select configurations for labelling. The configuration is selected if only $ \sigma_{lo} \leq \epsilon \leq \sigma_{up}$. In our study we use $\sigma_{lo} = 0.05$, $\sigma_{up} = 0.15$. After selecting all candidate structures, a fixed number (100 in our case) of them with the highest values of $\epsilon$ are used for labelling.

During the next (labelling) stage, for each atomic configuration selected during the exploration stage, reference values of energy, forces and virial tensors are calculated by density functional theory (DFT) as implemented in Vienna \textit{ab~initio} simulation program (VASP)~\cite{VASP1}. Projector augmented-wave (PAW) pseudopotentials and Perdew-Burke-Ernzerhof (PBE)~\cite{Perdew1992PRB1,Perdew1992PRB2} gradient approximation to the exchange-correlation functional~\cite{Kresse1999PRB} were applied.  To sample the Brillouin zone only the $\Gamma$ point was used. Energy cutoff and other related parameters was set in accordance to built-in precision parameter PREC=Normal.

The described above stages was repeated until the selection procedure is saturated that means the inequality $\epsilon \leq \sigma_{lo}$ is satisfied for all the configurations generated at the exploration stage. The resulting dataset includes about 10,000 atomic configurations and corresponding values of energies, forces and virial tensors calculated by DFT. 

\subsection{Verification of DP}

\begin{figure*}
	\begin{center}
		\includegraphics[width=0.33\linewidth]{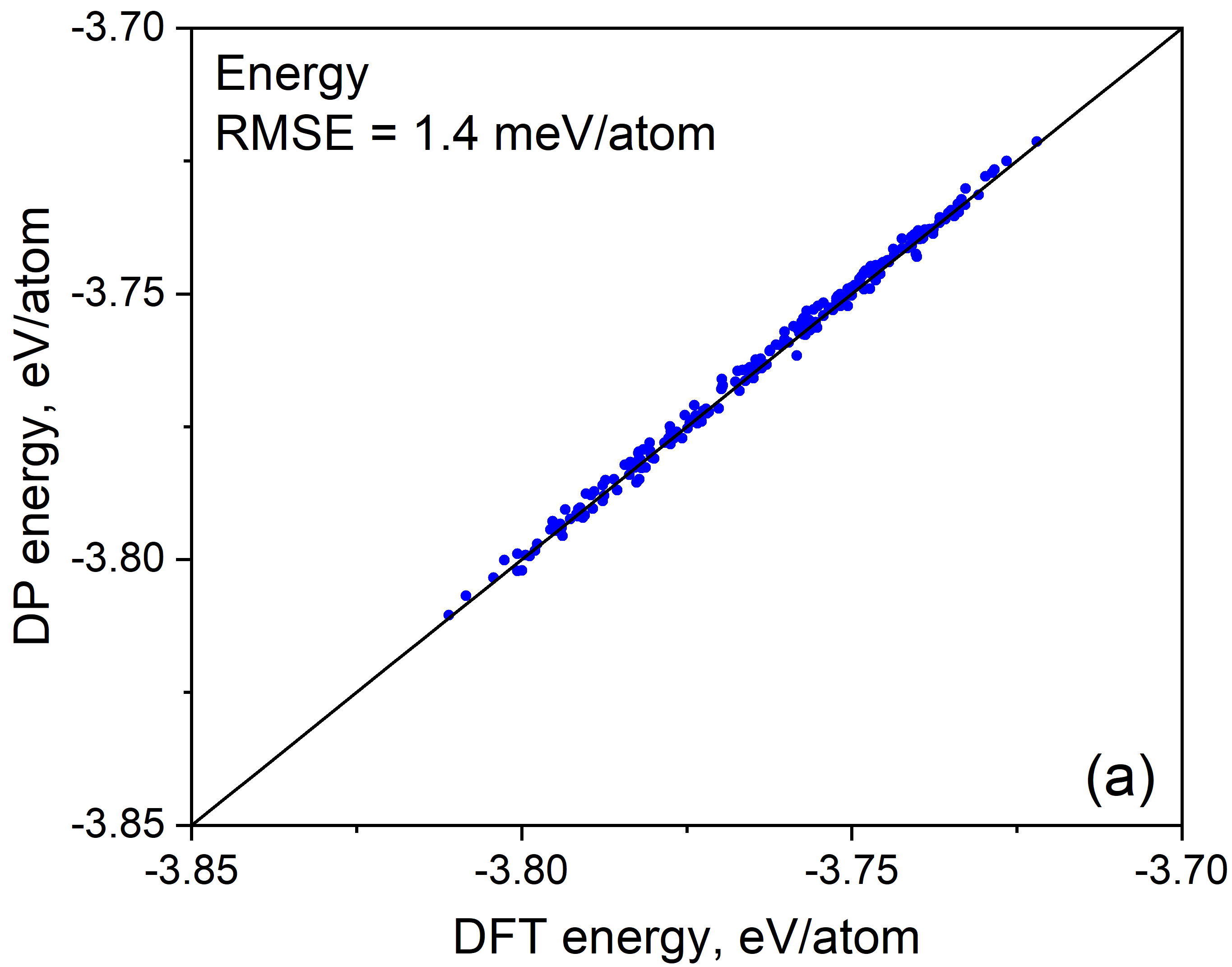}
		\includegraphics[width=0.31\linewidth]{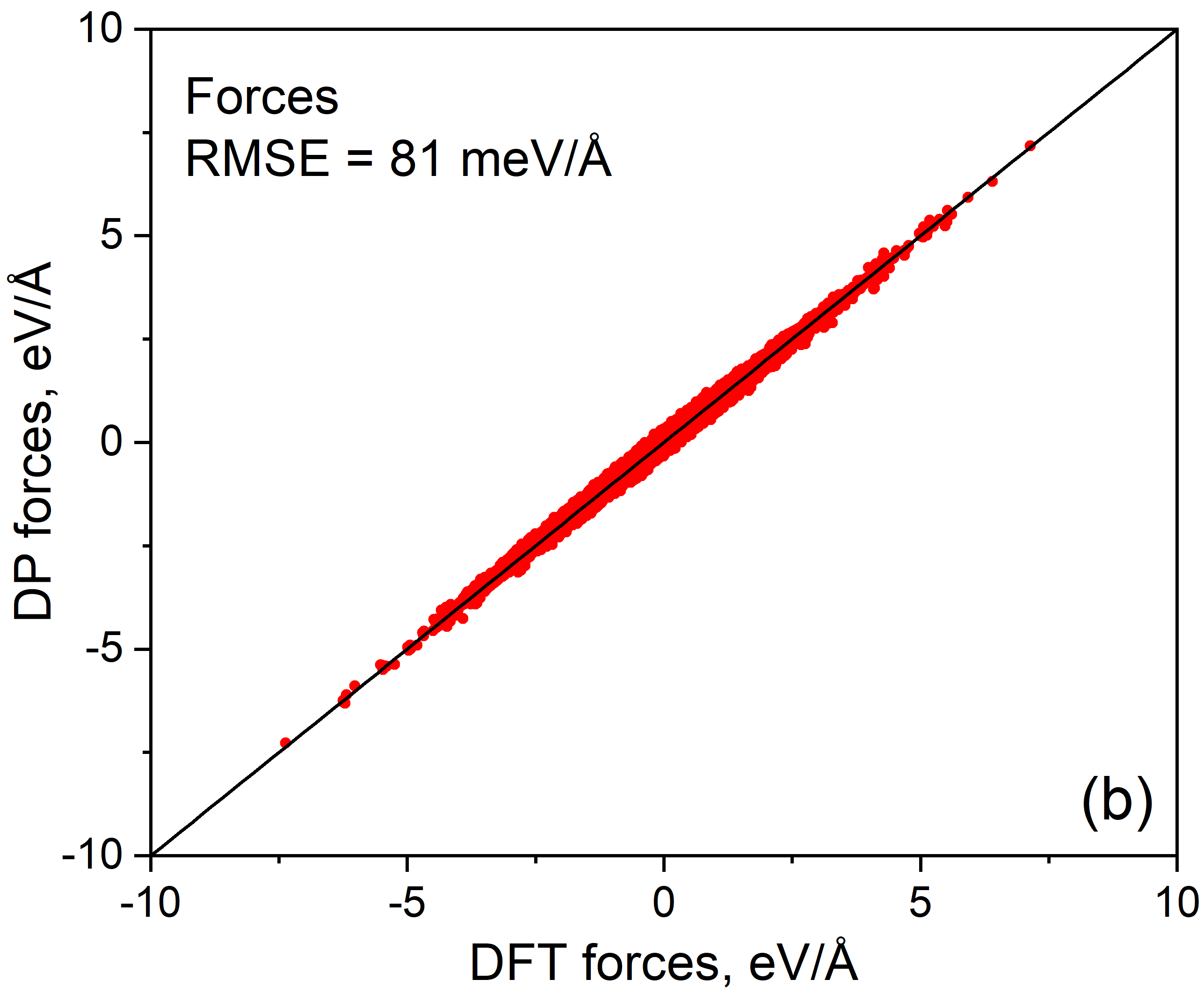}
		\includegraphics[width=0.31\linewidth]{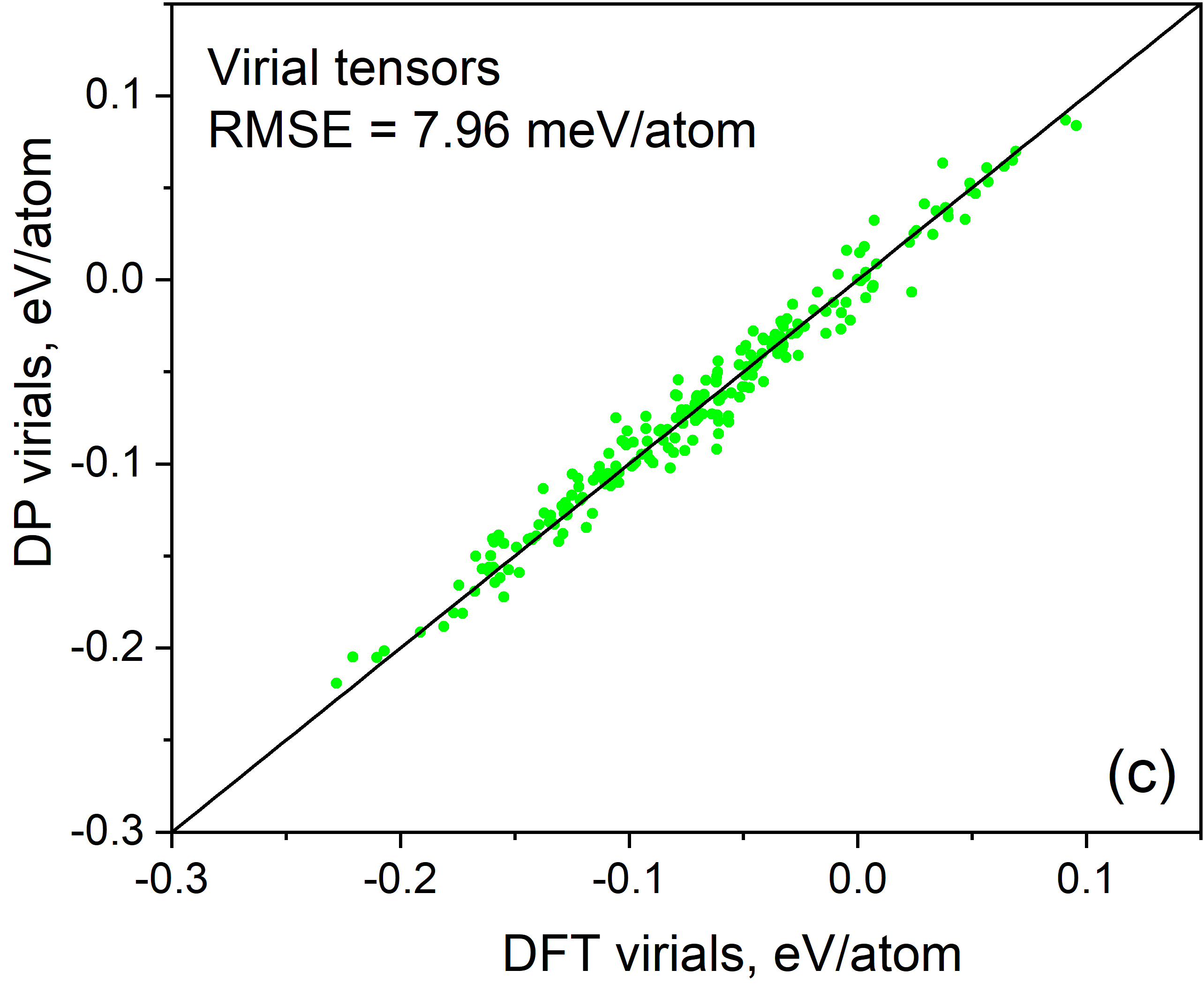}
		\caption{DP-based MD simulation results vs obtained using DFT for: (a) energies, (b) forces and  (c) virial tensors for Al-Cu-Ni melts.}
		\label{fig:dptest}
	\end{center}
\end{figure*}

\begin{figure*}
	\begin{center}
		\includegraphics[width=0.95\linewidth]{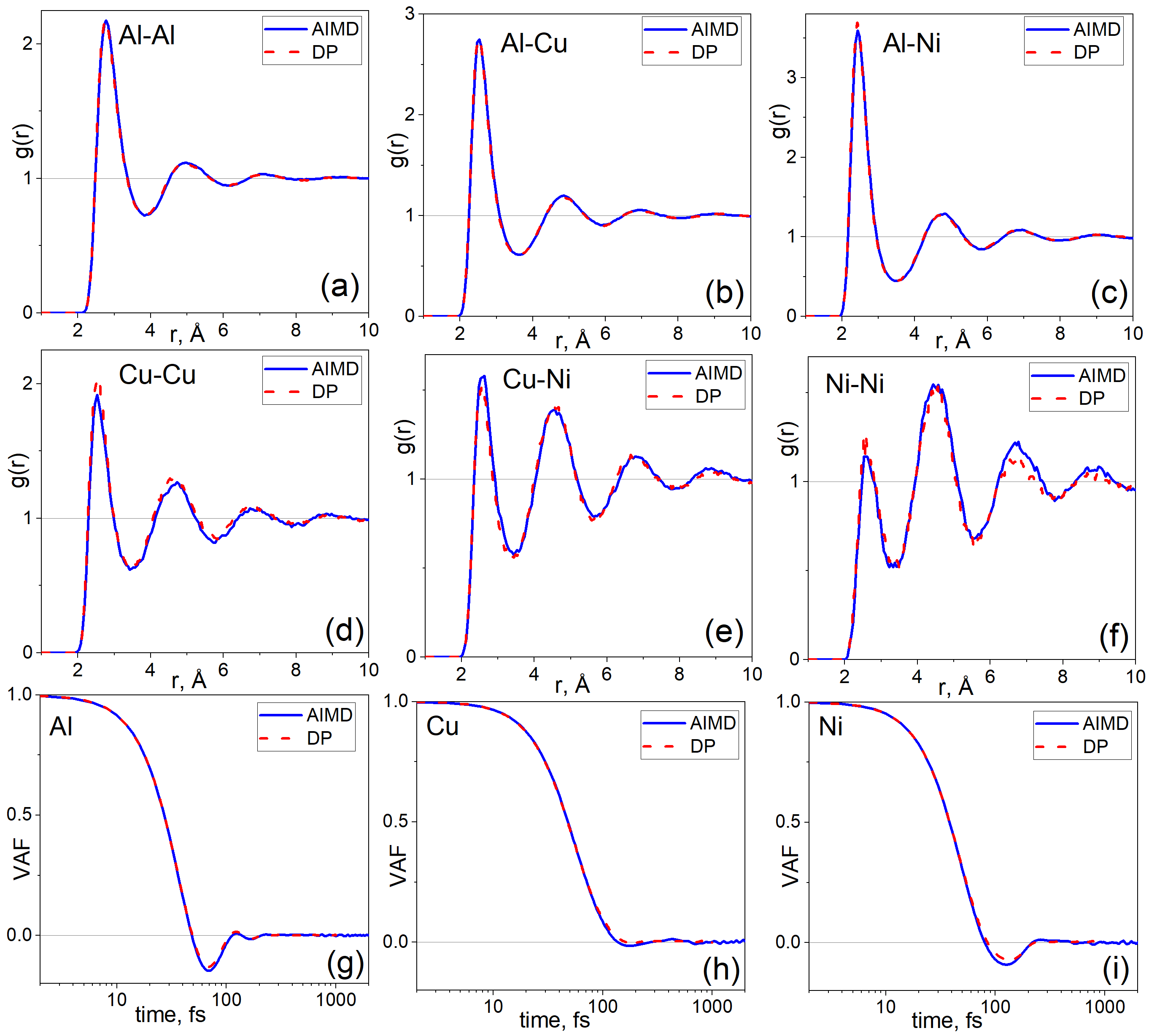}
		\caption{Partial radial distribution functions (a-f) and velocity autocorrelation functions (g-i) for ${\rm Al_{72.5}Cu_{17.5}Ni_{10}}$ melt at $T=1673$ K extracted from AIMD simulations (solid blue lines) as well as from DP simulations (red dashed lines).}
		\label{fig:rdf_vaf}
	\end{center}
\end{figure*}

Before calculating the viscosity, we verify the predictability of the developed DP in comparison to \textit{ab~initio} data. A standard first test is to check how the potential reproduces DFT values of energies, forces and virial tensors. In Fig.~\ref{fig:dptest} we show DP vs DFT correlations for these quantities calculated on a validation dataset including data from all compositions. One can find a pronounced linear correlation between presented data. The values of RMSEs for energy, forces, and virials are respectively 1.4 meV/atom, 81 meV/\r{A} and 7.96 meV/atom. So we expect that the potential would provide a good accuracy in simulations of the observable properties of Al-Cu-Ni melts at least for Al$_{100-x}$Cu$_{x}$Ni$_{10}$ compositions. To check this point, we have extracted partial radial distribution functions (RDFs) and velocity autocorrelation functions (VAFs) of the melts from both \textit{ab~initio} molecular dynamics (AIMD) and DP simulations and obtained close agreement between these data. In Fig.~\ref{fig:rdf_vaf} we illustrate this result with the example of ${\rm Al_{72.5}Cu_{17.5}Ni_{10}}$ melt at $T=1673$ K. Thus, we conclude that developed potential reproduce structural and dynamic properties of Al-Cu-Ni melts with {\it ab~initio} accuracy.

\section{Viscosity calculations}

The shear viscosity $\eta$ of melt can be obtained using both equilibrium (Green--Kubo) (G--K) and reverse non-equilibrium (M\"{u}ller-Plathe) (M--P) methods, here we use them for the verification of each other. The kinematic viscosity $\nu$ is obtained as $\eta / \rho$, where $\rho$ is the equilibrium density at a given tmeprature $T$.

\subsection{Non-equilibrium method}
\label{31}
The viscosity of liquids and fluids can be calculated by the reverse non-equilibrium molecular dynamics (RNEMD) approach \cite{Plathe1999}. It imposes momentum transfer in the simulation cell. As a response, a velocity gradient is established. The ratio of the momentum flux to the velocity gradient gives the shear viscosity:
\begin{equation}
\eta = \frac{j_z (p_x)}{\partial v_x / \partial z}
\end{equation}
\noindent where $j_z (p_x)$ is the flux of the $x$-component of momentum, and $\partial v_x / \partial z$ is the $x$-component of velocity gradient, both of them in $z$ direction.

To create the flow in the melts, we use \texttt{fix viscosity} command implemented in LAMMPS~\cite{lammps1}. The cell is divided into 50 bins in Z direction, and the average velocity of the group of atoms in each layer is calculated. Then, the velocities of Al atoms in the bottom and middle bins at each N step are swapped. As a result of such a procedure, the Couette flow is established in the unit cell.

\subsection{Equilibrium method}

In our previous papers, we have shown that the non-equilibrium simulation results converge to the Green--Kubo (G--K) values as the strain rate or the corresponding momentum flux decreases~\cite{Kondratyuk2019ind,Kondratyuk2019solo}. Thus, it is reasonable to use Green--Kubo technique because it gives zero shear rate values of viscosity. 

The Green--Kubo formula for the shear viscosity $\eta_{\alpha \beta}$ in $\alpha\beta$-plane is~\cite{Green1954,Kubo1957}:
\begin{equation}\label{eq:GK}
\eta_{\alpha \beta} = \lim_{t^{'}\rightarrow\infty} \dfrac{V}{k_{B}T}\int \limits_{0}^{t^{'}} C_{\sigma}(t) dt,
\end{equation}
\begin{equation}\label{eq:AC}
C_{\sigma} (t) = \langle \sigma_{\alpha \beta}(0)\sigma_{\alpha \beta}(t) \rangle,
\end{equation}
where $C_{\sigma} (t)$ is a shear-stress autocorrelation function (SACF), which averaging over the canonical ensemble is denoted by $\langle...\rangle$, $\sigma_{\alpha \beta}$ are shear components of the stress tensor, $V$ and $T$ are respectively system volume and temperature, and $k_{B}$ is the Boltzmann's constant. In practice, the integral in Eq.~\eqref{eq:GK} is typically calculated up to some finite time $t^{'}$ when $C_{\sigma}$ decays to zero within the accuracy of numerical simulation. The shear viscosity $\eta$ can be found as an average of $\eta_{xy}$, $\eta_{xz}$ and $\eta_{yz}$.

The stress tensor $\sigma_{\alpha \beta}$ is calculated from the following equation:
\begin{equation}\label{eq:Stress}
\sigma_{\alpha \beta}V = \sum \limits_{i=1}^{N} m_{i}v_{i_\alpha}v_{i_\beta} + \sum \limits_{i=1}^{N'}r_{i_\alpha}f_{i_\beta},
\end{equation}
where $N$ is a number of atoms, $N'$ includes atoms from neighboring sub-domains, $r_{i_\alpha}$ and $v_{i_\alpha}$ are $\alpha$-components of coordinate and velocity of the $i$-th atom, and $f_{i_\alpha}$ is $\alpha$-component of the force that acts on the $i$-th atom.

For effective averaging of G--K integrals, we use the time decomposition method (TDM) proposed by Maginn and co-authors~\cite{Zhang2015}. In this method, the G--K integral was averaged over a number of independent MD runs and then should be fitted by the double exponential function
\begin{equation}\label{eq:GK_fit}
\eta (t) = A \cdot \alpha \cdot \tau_{1} \cdot (1 - e^{-t/\tau_{1}}) + A \cdot (1-\alpha) \cdot \tau_{2} \cdot (1 - e^{-t/\tau_{2}}),
\end{equation}
where $A$, $\alpha$, $\tau_1$ and $\tau_2$ are the fitting parameters. The SACFs are calculated in LAMMPS during the simulations via the \texttt{fix ave/correlate} command. 

\subsection{MD Simulation details}

\begin{figure}[!ht]
	\begin{center}
		\includegraphics[width=0.99\linewidth]{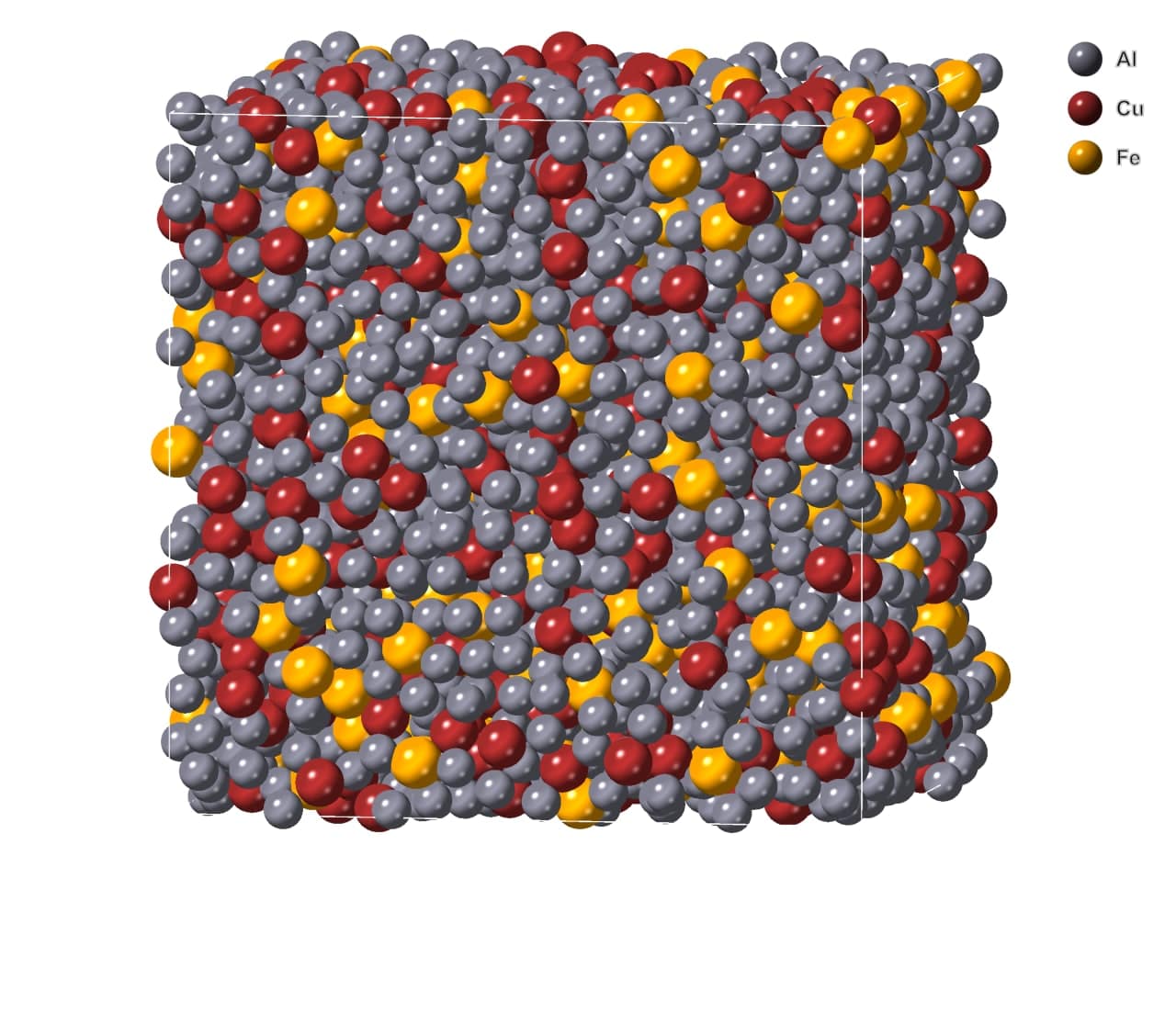}
		\caption{The snapshot of Al$_{72.5}$Cu$_{17.5}$Ni$_{10}$ melt with 4096 particles in the cell (Al – silver, Cu – red, and Ni – golden) used in the DP simulations with DP potential in LAMMPS.}
		\label{fig:unitcell}
	\end{center}
\end{figure}

Initial configurations for all Al$_{100-x}$Cu$_{x}$Ni$_{10}$ compositions are taken from the corresponding DFT simulations on which the DP was built. These configurations consist of 512 atoms with different compositions of Al, Cu and Ni atoms. We create 2x2x2 supercells of these configurations and carry out density equilibrations for 50~ps in the isothermal-isobaric ensemble (NPT) using the Nos\'{e}-Hoover barostat/thermostat~\cite{Nose,Hoover,Shinoda2004} with 1~fs integration timestep. After the NPT relaxation, we compress the unit cells to the average values of densities for 10~ps. The example of the unit cell for Al$_{72.5}$Cu$_{17.5}$Ni$_{10}$ melt is shown in Fig.~\ref{fig:unitcell}.

For the Green--Kubo calculations, we use about 100 trajectories of 50~ps length in the canonical ensemble to produce averaged SACFs. This length of a single trajectory is found to be enough for the statistical independence between each SACF, based on the diffusion coefficients of atoms. For RNEMD simulations, we carried out 1~ns-long trajectories to produce the velocity profiles. The first 50 ps were excluded from the consideration because of flow establishment.
 
\section{Viscosity of Al-Cu-Ni melts}

\subsection{Verification of methods}

To verify the kinematic viscosity calculation approach, we carry out both equilibrium Green--Kubo (G--K) and non-equilibrium M\"{u}ller--Plathe (M--P) simulations for Al$_{66}$Cu$_{24}$Ni$_{10}$ melt at 1673~K.

\begin{figure}[!ht]
	\begin{center}
		\includegraphics[width=0.9\linewidth]{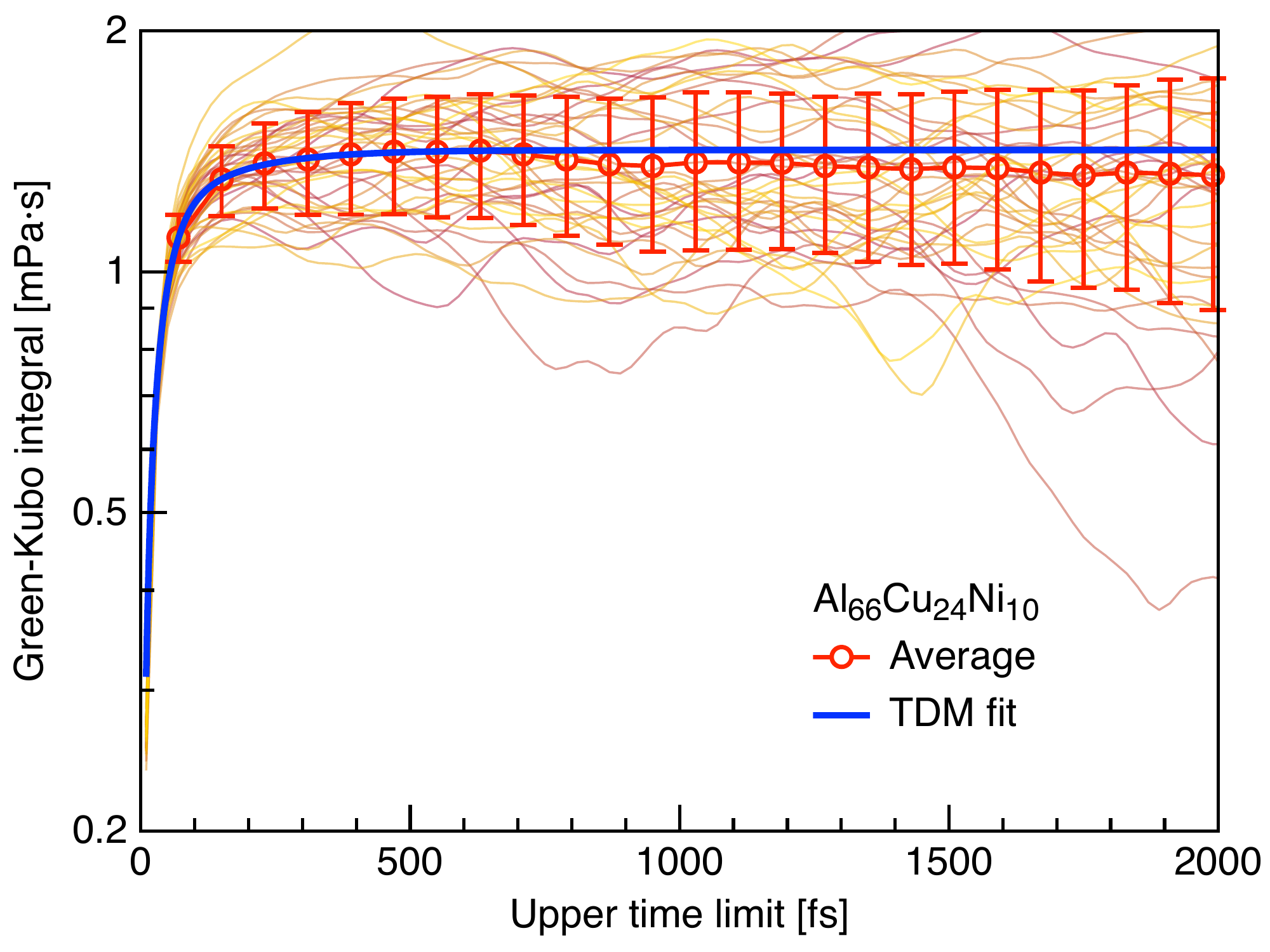}
		\caption{The dependence of Green--Kubo integral on the upper time limit for Al$_{66}$Cu$_{24}$Ni$_{10}$ melt at T=1673~K, $\rho=3.741$~g/cm$^3$.}
		\label{fig:gk}
	\end{center}
\end{figure}

The dependence of G--K integral~\eqref{eq:GK} on the upper time limit is shown in Fig.~\ref{fig:gk}. Solid lines represent the results for statistically independent MD trajectories. The average values of G--K integral at each upper time limit are shown with red points. One can find, while separate trajectories reveal large fluctuations of G--K integral,  the average value converges to a well defined value at about 1~ps. We approximate the averaged values with the double exponential function Eq.~\eqref{eq:GK_fit} treating the points with the weight $\sigma^{-1/2}$, where $\sigma$ is a standard deviation calculated over independent MD trajectories. The final value of kinematic viscosity $\nu$ was found as the infinite time limit of the approximation Eq.~(\ref{eq:GK_fit}) divided by the equilibrium density $\rho$ at given temperature $T=1673$~K which is equal to 3.741~g/cm$^3$ for Al$_{66}$Cu$_{24}$Ni$_{10}$. Thus, the resulting value is $(3.9\pm 0.2)$$\cdot10^{-7}$m$^{2}$/s.

\begin{figure}
\centering
   (a)\includegraphics[width=0.95\linewidth]{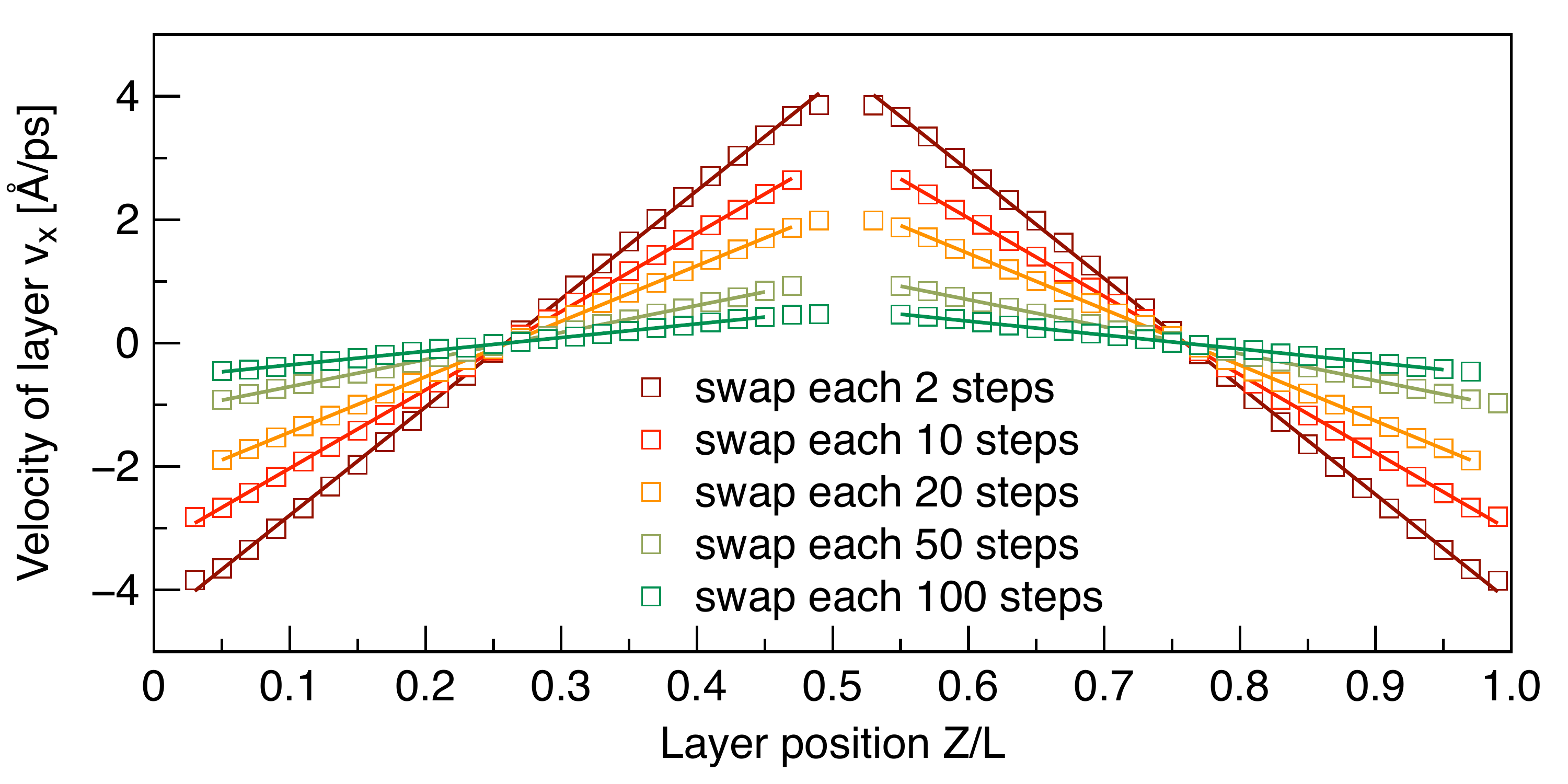} \\ 
   (b)\includegraphics[width=0.95\linewidth]{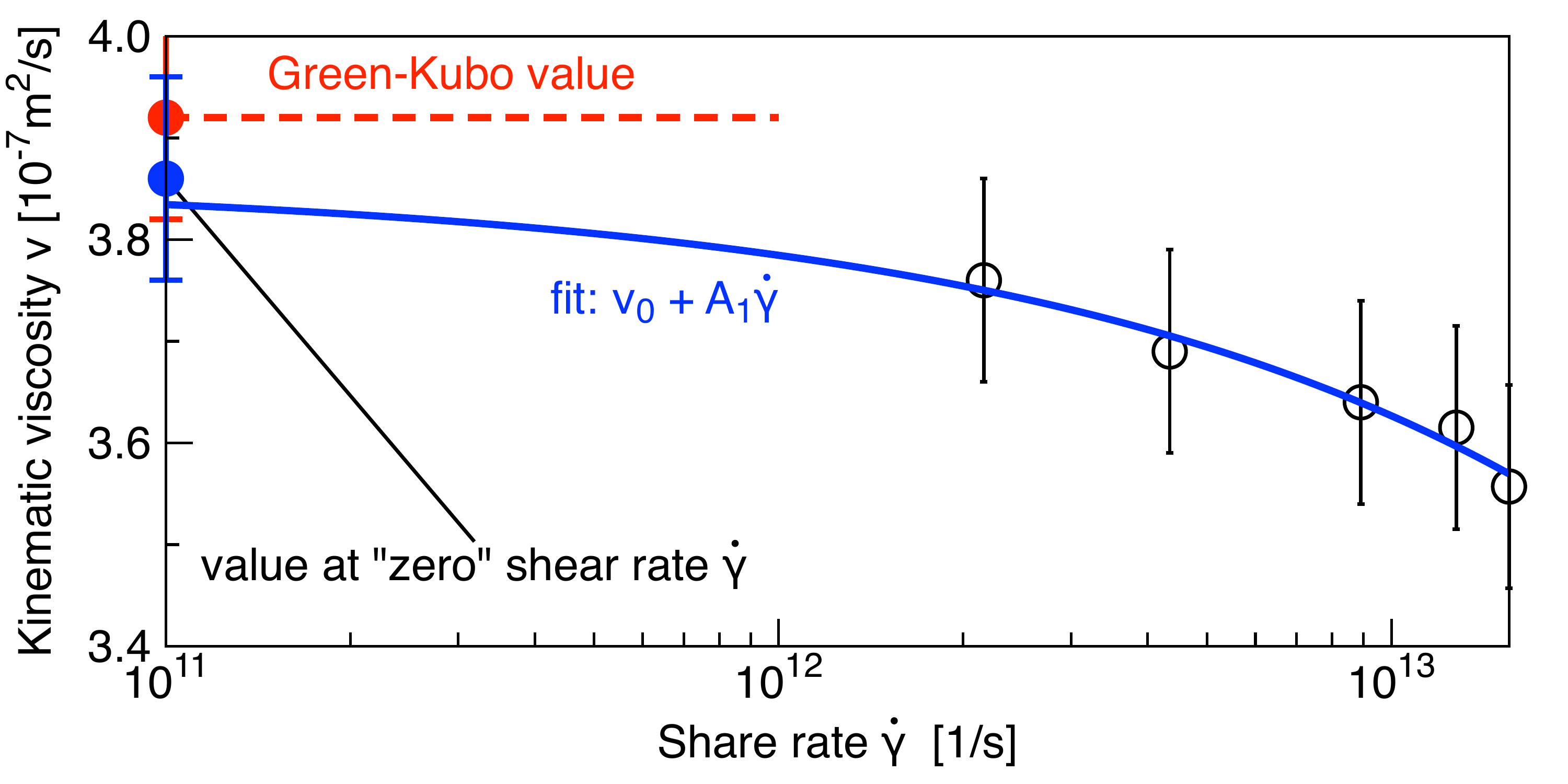} \\ 
   \caption{(a) The velocity profiles obtained via RNEMD simulation for Al$_{66}$Cu$_{24}$Ni$_{10}$ melt at $T = 1673$ K, $\rho=3.741$~g/cm$^3$ for the different values of momentum fluxes. (b) The dependence of the kinematic viscosity on the shear rate. The red point shows the equilibrium value of viscosity calculated via the Green--Kubo method. The blue one is the value at $\dot{\gamma}=0$, found from~Eq.\eqref{eq:shear}.}\label{fig:nemd}
\end{figure}

Moreover  we studied  the kinematic viscosity dependence on the strain rate in the M--P method. For this purpose, the swapping frequency of Al atoms velocities in bottom and middle bins is varied, which establishes different values of strain rates (See Sec.~\ref{31} for details). The obtained velocity profiles are shown in Fig.~\ref{fig:nemd}a for different values of swap rates. The profiles are linear, which corresponds to Couette flow regime. 

The dependence of the kinematic viscosity on the shear rate $\dot{\gamma}$ is shown in Fig.~\ref{fig:nemd}b. One can see, for the considered shear rates the kinematic viscosity grows monotonically  as the $\dot{\gamma}$ decreases.  Such non-Newtonian response is an expected behaviour for atomic~\cite{Han2008} and molecular~\cite{Kondratyuk2019ind} liquids at the shear rates achievable in MD. For liquid copper, Han et al.~\cite{Han2008} observed similar non-Newtonian dependence $\eta(\dot{\gamma})$ via RNEMD and used the following approximation equation to calculate zero-shear viscosity $\eta_0$:
\begin{equation}\label{eq:shear}
    \eta = \eta_{0} + A_{1}\dot{\gamma}^{1/2}.
\end{equation}
 This equation is a consequence of Mode Coupling Theory~\cite{MCT}. 
 We use the same equation to fit obtained $\nu(\dot{\gamma})$ dependence for Al$_{66}$Cu$_{24}$Ni$_{10}$ melt. The resulting approximation curve is presented by solid blue line in \ref{fig:nemd}b. Excellent agreement with the RNEMD results is achieved. The value of zero shear rate kinematic viscosity from this approximation is $(3.85\pm 0.2)$$\cdot10^{-7}$m$^{2}$/s.
 
 It was suggested, that for ionic liquids, there are two regimes for $\eta(\dot{\gamma)}$ could appear, this caused as a consequence of different relaxation times of internal molecular degrees of freedom~\cite{Oanh2010,Safinejad2018}. As one can see from Fig.~\ref{fig:nemd}b, here only a single regime is observed that is rather expected from equilibrium metallic melts.

Thus, we conclude that the kinematic viscosity coefficients obtained by both equilibrium (G--K) and non-equilibrium (M--P) methods agree within the accuracy of the methods. That confirms that the predicting power of MD simulations of the viscosity is determined only by the developed potential. In the following, the Green--Kubo method is used to calculate the zero-shear kinematic viscosity in Al-Cu-Ni melts.

\subsection{Concentration dependence of the viscosity}

A challenging task for any theoretical method is a calculation of the concentration dependencies of the viscosity in metallic melts. Such dependencies often demonstrate a non-monotonic behavior with extrema located in the vicinity of phase boundaries on phase diagrams~\cite{Kamaeva2020,TAN2007PhysB,Beltyukov2010,Ladyanov2014PhysChemLiq}. The unequivocal explanation of such behavior is not always available but it is obviously caused by a change in chemical interaction and local structure of melts during  the concentration change. The simulation of such non-linear behaviour of the viscosity change is a complicated problem and  can be treated as a stress-test for machine learning potentials.

\begin{figure}[!ht]
	\begin{center}
		\includegraphics[width=0.9\linewidth]{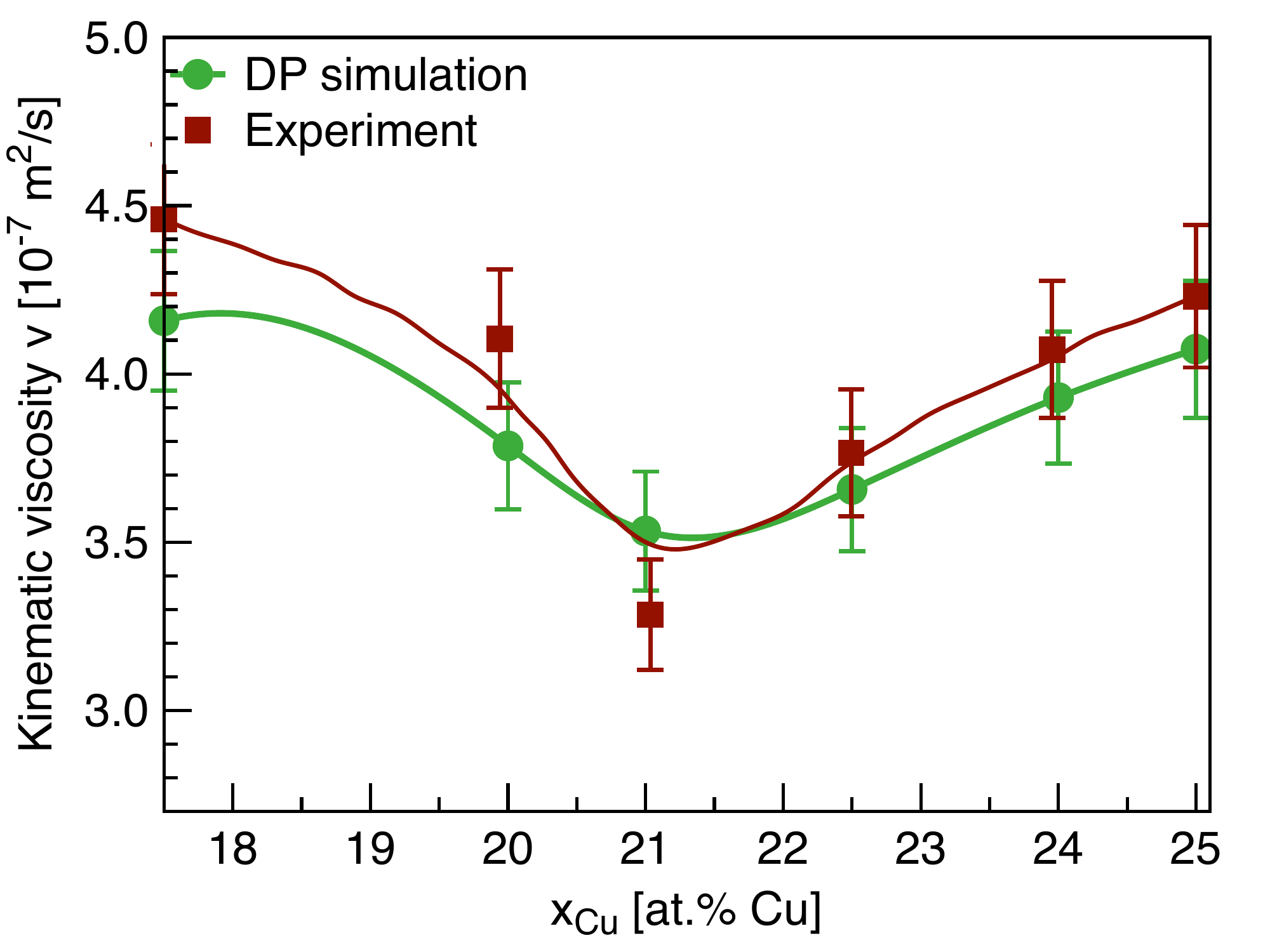}
		\caption{The  kinematic shear viscosity $\nu$ dependence on the $x_\mathrm{Cu}$ at.\% concentration for Al$_{100-x}$Cu$_{x}$Ni$_{10}$ melt at T=1673~K. Green points are the results of the current work (with curve in order to guide the eye). Red points are the experimental data from~\cite{Kamaeva2020} with the approximation curve.}
		\label{fig:visco_x}
	\end{center}
\end{figure}

\begin{figure}[!ht]
	\begin{center}
		\includegraphics[width=0.9\linewidth]{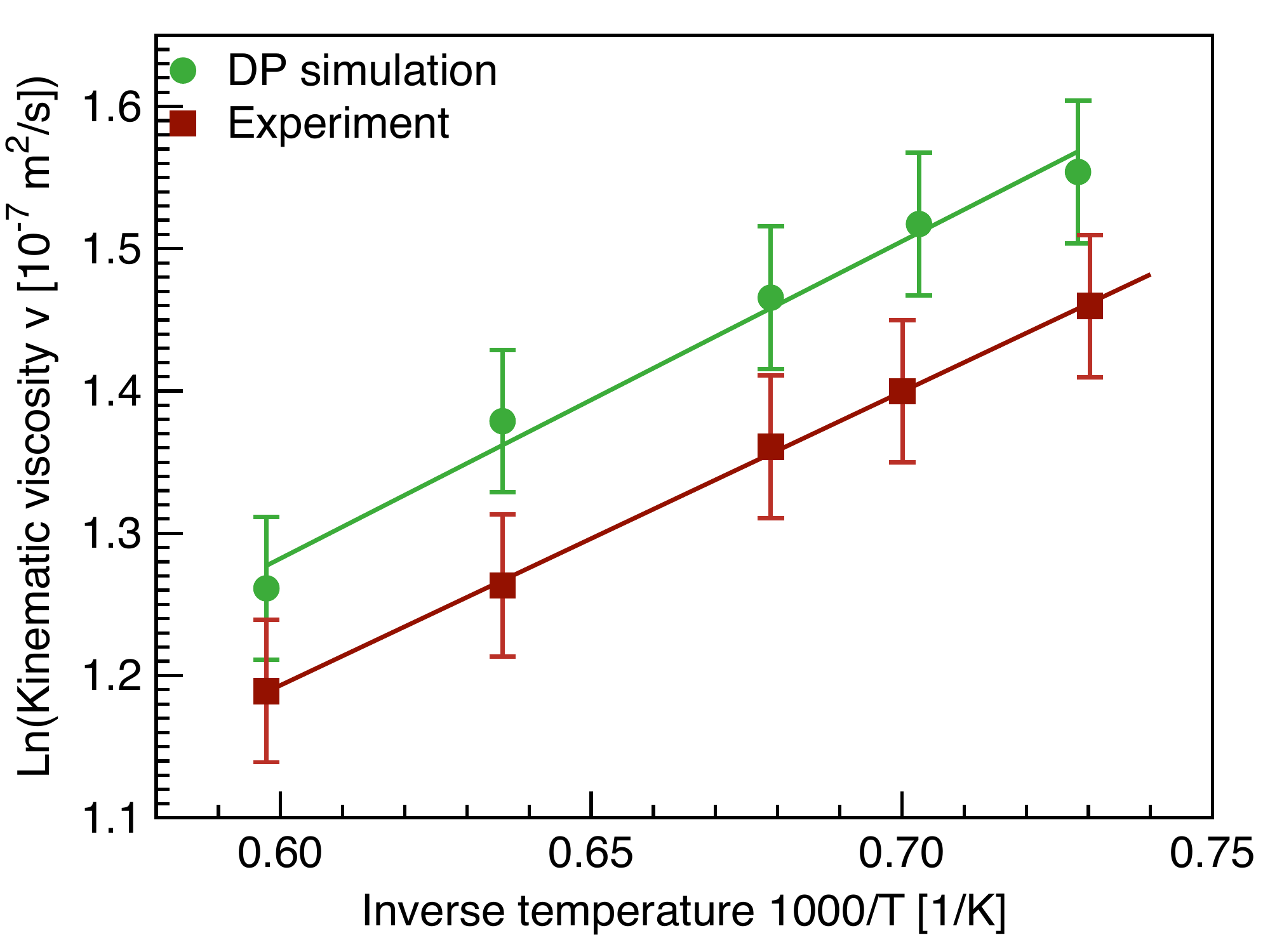}
		\caption{The dependence of log of kinematic shear viscosity $ln(\nu)$ for Al$_{69}$Cu$_{21}$Ni$_{10}$ melt on the inverse temperature $1000/T$. Green and red bullets represent respectively simulated and experimental~\cite{Kamaeva2020} data. }
		\label{fig:visco_T}
	\end{center}
\end{figure}

In 2017, Mudry et al.~\cite{Mudry2017} studied the influence of Ni additions on viscosity of Al$_{2}$Cu melt. They demonstrated the viscosity decrease with the Ni addition due to the structural changes in the liquid. Recently, we  experimentally observed that the kinematic viscosity in Al$_{100-x}$Cu$_{x}$Ni$_{10}$  melts experiences a minimum at $x_{\rm Cu} = 21$~at\% which corresponds to the eutectic point~\cite{Kamaeva2020}. To check whether this behaviour is reproduced by DP-based MD simulation, we calculated the kinematic viscosity coefficients along correspondent concentration cross-section at $T = 1673$~K varying $x_{\rm Cu}$ in the range of 19-25~at\%. The comparison of experimental data from~\cite{Kamaeva2020} and simulation results obtained in present work are presented in Fig.~\ref{fig:visco_x}. One can see that DP-simulations provide the quantitative agreement with experimental data; the simulated $\nu (x)$ curve reproduces the minimum at eutectic point; the average deviation from experiment is about 8~\% and the maximum deviation is 9~\%. Such results should be considered as very accurate, taking into account the complexity of the problem.

\subsection{Temperature dependence of the viscosity}

To test whether the developed potential reproduces the temperature dependence of viscosity, we carry out the G--K calculations for Al$_{69}$Cu$_{21}$Ni$_{10}$ melt at different temperatures. The temperature range was 1373--1673~K, in which the system remains liquid according to both experimental results and our simulations. Before the viscosity calculations, NPT relaxations for 50~ps are carried out to determine equilibrium densities. Then, we deform the unit cells to the corresponding target densities. The comparison of the calculated temperature dependency of kinematic viscosity with the experimental data is presented in Fig.~\ref{fig:visco_T}. One can see that $\ln\nu$ vs $1/T$ curves for experimental and simulated data are the linear slopes and both have very close slope angles.  Thus, the Arrhenius equation can be used to approximate the obtained polytherms, which assumes the momentum transport as an activated rate process with activation energy $E_{a}$:

\begin{equation}\label{eq:arrh}
    \nu = \nu_{0} \exp{(E_{a}/k_{B}T)},
\end{equation}
where $\nu_{0}$ is a is the pre-exponential constant  under investigation. It is widely accepted that Arrhenius equation is a good approximation for many metals and metallic alloys~\cite{Rajendra1990,Konstantinova2009,Olyanina2016}.

The Arrhenius fit gives $\nu_{0}$=9.57$\cdot$10$^{-8}$~m$^{2}$/s, $E_{a}$ = 17.15~kJ/mol for the experimental data and $\nu_{0}$=9.47$\cdot$10$^{-8}$~m$^{2}$/s, $E_{a}$ = 18.53~kJ/mol for our simulations data. Thus, the developed DP almost reproduces the pre-factor $\nu_{0}$ and slightly underestimates the activation barrier $E_{a}$.



\section{DISCUSSION AND CONCLUSIONS}

In the present paper we address the question of if one can accurately calculate the viscosity in multicomponent metallic melts. As we mentioned above, neither traditional classical molecular dynamics (MD) (based on empirical potentials like embedded-atom method [EAM]  potentials) nor \textit{ab~initio} can not be considered as an effective and universal tool for solving this problem. The former approach usually suffers from lack of accuracy and the latter is too time-consuming and so usually cannot provide enough statistics. A possible solution is the use of machine learning potentials which demonstrate optimal accuracy/efficiency ratio.  Here we make a first step to check this idea.

Considering deep neural network potentials (DP) as a regression model for machine learning potentials and DeePMD-kit as a powerful tool for their development, we calculate both concentration and temperature dependencies of kinematic viscosity in Al-Cu-Ni melts. The main results are presented in Figs.~\ref{fig:visco_x}, \ref{fig:visco_T}. Analysing the results, we conclude that the developed DP allows one to calculate viscosity with high accuracy; the deviation from experimental data does not exceed 9\% and is close to the uncertainty interval of experimental data. More importantly, DP-based MD simulations reproduce the minimum on concentration dependency of the viscosity at the eutectic point. Thus, we response that for the Al-Cu-Ni melts one can positively answer to the main question raised in the title of the paper.

Despite the general success in calculating of the viscosity of Al-Cu-Ni melts, the agreement between theory and experiment is not perfect. Now we discuss briefly the possible reasons for such deviation from experimental data and general ways to improve the results.

First, it should be noted that measurement of the viscosity of metallic melts is a complicated experimental procedure especially in the case of alloys with high liquidus temperatures (> 1000 K)~\cite{Brooks2005, Cheng2014}. As a result, experimental data obtained for the same system by different methods and/or on different devices can differ from each other by several times (see, for example,~\cite{Dinsdale2004,Chen2014PhilMag} for Al and its alloys). We use experimental data obtained with an oscillating vessel viscometer using precise and carefully approved procedure~\cite{Beltyukov2008}. However, some systematic errors caused by the individual characteristics of the experimental device are possible here. Unfortunately, for the best of our knowledge, we do not know the alternative data for the viscosity of Al-Cu-Ni melts and so can not fully get rid of  this error.

 Besides experimental difficulties discussed above, there are many sources of uncertainties in calculating shear viscosity via MD simulations. The main ones are: (i) statistical errors, (ii) finite-size effects, and (iii) imperfections of interatomic potentials. Recently, Kim et al.~\cite{Kim2018JCP} carefully analyzed the first two of them. Based on their findings and taking into account the results of the method verification presented above, we can declare that our calculations are not noticeably affected by the finite size effects and by the lack of statistics. Thus, the main source of uncertainties in our simulations can come from possible drawbacks of the potential.

Again, there are many reasons that can cause inaccuracies in machine learning interatomic potentials, such as insufficient training dataset, poor functional form of the potential and errors in training procedure. We carefully verify our DPs and conclude that it provides \textit{ab~initio} accuracy for interatomic energies, forces and virial tensors. Thus, the only noticeable source of DP imperfection is an inaccuracy in \textit{ab~initio} data itself. This inaccuracy may be noticeable when one applies standard DFT calculations for transition metals (see, for example, our recent results for pure Ni~\cite{Ryltsev2022JMolLiq}). We suppose that this problem can be even more serious when considering melts containing four and more elements including transition metals. A possible solution in this situation is going beyond DFT to generate training dataset. This is an urgent challenging task, whose applicability extends well beyond the problem of viscosity calculation.

Anyway, here we declare that the accuracy achieved in this paper for ternary metallic melt is high enough and so ML-based MD simulations is highly promising way to calculate viscosity in multicomponent metallic melts. 
Moreover in this perspective, dare we hope not only for reproduction of the experimental results but that one can apply the described procedure on the prediction of the melts properties. 

\section*{ACKNOWLEDGEMENTS}

We thank L. Kamaeva for helpful discussions of experimental data end techniques. This work was supported by the Russian Science Foundation (grant 22-22-00506). N. Kondratyuk was supported by the strategic academic leadership program ``Priority 2030'' (Agreement  075-02-2021-1316 30.09.2021). The numerical calculations are carried out using computing resources of the federal collective usage center 'Complex for Simulation and Data Processing for Mega-science Facilities' at NRC 'Kurchatov Institute' (ckp.nrcki.ru), supercomputers at Joint Supercomputer Center of Russian Academy of Sciences (www.jscc.ru), 'Uran' supercomputer of IMM UB RAS (parallel.uran.ru) and HPC facilities at HSE University~\cite{Kostenetskiy2021}.

\section*{Data Availability Statement}

The data that supports the findings of this study is available from N.K. upon reasonable request.

 \bibliographystyle{elsarticle-num} 
 \bibliography{cas-refs}





\end{document}